\documentclass[runningheads]{llncs}
\usepackage{graphicx}
\usepackage{amsmath}
\usepackage{amssymb}
\usepackage{amsfonts}
\usepackage{verbatim}
\usepackage{float}
\usepackage[symbol]{footmisc}

\renewcommand{\thefootnote}{\fnsymbol{footnote}}

\PassOptionsToPackage{hyphens}{url}\usepackage[colorlinks]{hyperref}

\makeatletter
\newcommand\footnoteref[1]{\protected@xdef\@thefnmark{\ref{#1}}\@footnotemark}
\makeatother

\begin{document}
\title{Improving Performance Estimation for FPGA-based Accelerators for Convolutional Neural Networks}
\titlerunning{Improving Performance Estimation for FPGA-based Accelerators}

\author{Martin Ferianc\inst{1,\footnotemark[4]}\and Hongxiang Fan\inst{2}\and Ringo S. W. Chu\inst{3} \and Jakub Stano\inst{4} \and Wayne Luk \inst{2}}

\authorrunning{M. Ferianc et al.}
%
\institute{Department of Electronic and Electrical Engineering, University College London, London, UK \\
\email{martin.ferianc.19@ucl.ac.uk}\and
Department of Computing, Imperial College London, London, UK
\\
\email{\{h.fan17, w.luk\}@imperial.ac.uk}
\and
Department of Computer Science, University College London, London, UK \email{ringo.chu.16@ucl.ac.uk}\\
\and
Department of Information Technology and Electrical Engineering, ETH Zurich, Zurich, Switzerland
\\
\email{jstano@ethz.ch}
}
\maketitle              
\begin{abstract}
Field-programmable gate array (FPGA) based accelerators are being widely used for acceleration of convolutional neural networks (CNNs) due to their potential in improving the performance and reconfigurability for specific application instances. To determine the optimal configuration of an FPGA-based accelerator, it is necessary to explore the design space and an accurate performance prediction plays an important role during the exploration. This work introduces a novel method for fast and accurate estimation of latency based on a Gaussian process parametrised by an analytic approximation and coupled with runtime data. The experiments conducted on three different CNNs on an FPGA-based accelerator on Intel Arria 10 GX 1150 demonstrated a {30.7\%} improvement in accuracy with respect to the mean absolute error in comparison to a standard analytic method in leave-one-out cross-validation.

\end{abstract}

\footnotetext{Corresponding author.}

\renewcommand*{\thefootnote}{\arabic{footnote}}
\setcounter{footnote}{0} 

\section{Introduction}
Field-programmable gate arrays (FPGAs) are becoming increasingly popular in the deep learning community, particularly in the acceleration of convolutional neural networks (CNNs) \cite{fan2018real,lian2019high,fan2019f}. This acceleration is achieved by parallelising the extensive concurrency exhibited by CNNs. As such, FPGA is ideal as the platform allows implementation of fine-grain parallelisations. To do an architectural exploration and determine the optimal hardware configuration, it is necessary to estimate the performance with respect to multiple different hardware specifications.

There are several performance estimation frameworks for reconfigurable FP\-GA-based accelerators \cite{yasudo2018performance,dai2018fast,enzler2000high},  
however, estimating the performance without knowing about scheduling is still a very challenging task because of two main reasons. First, the explicit time to execute a certain operation on hardware varies by on/off-chip communication, synchronisation, control signals, I/O interruptions and in particular for the CNN accelerators - the CNN's architecture, which complicate analytic estimation. Second, it is difficult to accurately select the most representative design features for all hardware specifications during the performance estimation.

In this paper, we propose a novel approach for performance estimation for FPGA-based CNN accelerators \cite{lian2019high}. This method constitutes a Gaussian process (GP) \cite{williams1996gaussian} coupled with a standard analytic method and statistical data. Gaussian process is a stochastic process, such that every finite collection of random variables has a multivariate normal distribution \cite{rasmussen2003gaussian}. Experiments were conducted on three different CNNs on Intel Arria GX 1150 FPGA and we compared the method to linear regression (LR), GP with zero mean function, GP with an artificial neural network (ANN) mean function \cite{gp_dnn}, gradient tree boosting (GTB) and ANN in estimating latency. We show that the proposed method achieved the top result among all compared methods.\footnote{\label{note1}A tutorial code is available at \url{https://git.io/Jv31c}.}

In Section \ref{sec:background} we demonstrate the standard approach for analytic performance estimation. Afterwards, in Section \ref{sec:gp} we introduce the proposed method, followed by Section \ref{sec:acc_data} where we describe the accelerator as well as the dataset on which we benchmarked the method. Then we present the evaluation in Section \ref{sec:eval} followed by a conclusion in Section \ref{sec:conclusion}.

\section{Background}\label{sec:background}
The most accurate method of determining the performance is escalating the CNN onto the hardware. One major drawback of this method is requiring re-synthesis and re-implementation for different hardware specifications. Therefore, it is more feasible and practical to perform the design space exploration (DSE) \cite{holland2011analytical} with respect to an estimate of the performance in a software level, rather than running the CNN each time for a different hardware configuration.

Even with a more advanced option of performance estimation, high irregularity within a complex accelerator results in case-by-case estimation. Therefore, this approach is unfeasible in general case, as it is usually constrained to a single hardware configuration. In our work, we are focused on estimating one particular aspect of performance - latency for a CNN reconfigurable accelerator. 

The standard 2D convolution layers, from which the CNN is constructed, occupy over 90\% of the overall processing time \cite{VenierisStylianos2018TfMC} and their latency $T_{i}$ on the accelerator needs to be estimated to determine the best hardware configuration through DSE. For 2D convolution, there are several categories of parallelism including filter parallelism ($PF$) or channel parallelism ($PC$) in addition to spatial and kernel parallelisms. These are the parameters that usually need to be determined during the DSE.

A performance estimation framework for reconfigurable dataflow platforms was proposed by Yasudo \textit{et al.} \cite{yasudo2018performance} that can analytically determine the number of accelerators suitable for the application. Dai \textit{et al.} \cite{dai2018fast} proposed an estimation method based on a GTB and a high-level synthesis report and they compared it with LR and ANN. However, their method requires a significant amount of data and features from the synthesis report, which might not be available, especially when high-level synthesis is not being used to describe the accelerator. Enzler \textit{et al.} proposed a general heuristic-based method \cite{enzler2000high} for estimating the performance of accelerator designs, which can be modified for CNN accelerators and is now used as the standard method.

\begin{table}
\centering
\caption{Notation used for performance estimation in an FPGA-based accelerator for convolutional neural networks.}
\label{tab:notation}
{
\begin{tabular}{|l|l|l|}\hline
\textbf{Parameter} & \textbf{Description}\\
\hline
$H$& Height of input feature map\\ \hline
$W$& Width of input feature map\\ \hline
$H_O$& Height of output feature map\\ \hline
$W_O$& Width of output feature map\\ \hline
$K$& Kernel size\\ \hline
$F$& Number of filters\\ \hline
$C$& Number of channels\\ \hline
$PF$& Parallelism in filter dimension\\ \hline
$PC$& Parallelism in channel dimension\\ \hline
$M_{CLK}$ [MHz] & Memory access clock cycle time \\ \hline 
$L_{CLK}$ [MHz] & Logic clock cycle time \\ \hline
$M_{EFF}$ [\%] & Memory transfer efficiency                     \\ \hline
$S$ [bits] & Memory transfer size \\ \hline 
$DW$ [bits] & Processing data width \\ \hline \hline 
$M$& Number of input features\\ \hline
$N$& Number of layers in a CNN\\ \hline
$P$& Number of training samples\\ \hline
\end{tabular}}{}
\end{table}

\begin{table}
\centering
\caption{Number of operations and a data size for a 2D convolution $i$.}
\label{tab:conv_operations}
\begin{tabular}{|l|l|}
\hline
   \textbf{Sizes}             & \textbf{Number of operations/Data size}   \\ \hline
   Number of compute operations  & $F_i \times C_i \times H_i \times W_i \times K_i \times K_i$                                  \\ \hline
   Input size&  $H_i \times W_i \times C_i$  \\ \hline
   Weights size & $F_i \times C_i \times K_i \times K_i$     \\ \hline 
   Output size &  $H_{O_i} \times W_{O_i} \times F_i$ \\ \hline
\end{tabular}
\end{table}

The simplest form of a heuristic for estimating latency on a hardware accelerator consists of dividing the overall processing time for a single input $T$ into time steps $T_{i}$ which correspond to the time to perform one 2D convolution in a feed-forward CNN consisting of $N$ 2D convolutions. The total estimated latency for the CNN in that given configuration is then simply added as $T = \sum_{i=1}^{N} T_{i}$. 

The time $T_{i}$ is being split into three different terms: (1) On-chip memory loading time $T_{load_i}$, (2) Computation time $T_{compute_i}$ and (3) Off-chip memory storing time $T_{store_i}$. Assuming the design is pipelined, the runtime $T_i$ is then decided by the slowest path which is chosen by the maximum among $T_{load_i}$, $T_{compute_i}$ and $T_{store_i}$. Each of these terms depends on a mixture of parameters that are specified by the 2D convolution: \textit{Input size, Output size, Number of compute operations}, device specific settings: \textit{Memory bandwidth, Clock cycle time} or the hardware architecture: \textit{Parallelism}, which are known prior to making a prediction. The estimated latency per layer is then computed as shown in Equations \ref{eq:heuristic} below
\begin{align}
    T_{load_{i}} &= \frac{\textrm{Input size}}{\textrm{Memory bandwidth}}\ \ \ \ \
    T_{compute_{i}} = \frac{\textrm{Number of compute operations}}{\textrm{Clock cycle time} \times \textrm{Parallelism}} \nonumber\\
    T_{store_{i}} &= \frac{\textrm{Output size}}{\textrm{Memory bandwidth}}\  \ \ \ \
    T_{i} = max\left(T_{load_{i}},T_{compute_{i}}, T_{store_{i}}\right)
    \label{eq:heuristic}
\end{align}

The heuristic approach does not depend on any statistical data to perform the estimation and it is simple to implement since it relies only on the features that can be easily read from the respective datasheets. Nevertheless, this general estimation method usually computes the most optimistic estimate and it does not leave room for delays caused by communication, synchronisation or control. One way to refine the estimation is that we can collect runtime data and use this data to improve the estimate. Therefore, in our work, we are proposing to use the standard analytic method as a mean function inside a GP together with the profiling data collected by running the CNN on real hardware to train the GP to model the observed misestimation.

\section{Gaussian Process with an Analytic Mean Function}\label{sec:gp}
GP is a modelling function built around Bayesian modelling which can embody our prior knowledge/model into our target \cite{rasmussen2003gaussian}. A GP is specified by a mean function $m(.)$ and a covariance function (kernel) $k(.,.)$. The mean function represents the supposed average of the estimated data. The kernel computes correlations between inputs and it encapsulates the structure of the hypothesised function. The main benefit of using a GP over other methods such as LR, GTB or ANN is that it can use the developed analytic foundations, such as the standard analytic performance estimation, as prior knowledge in a form of $m(.)$.

The predictive distribution of the GP, $p(\mathbf{y_{T}}|\mathbf{X},\mathbf{y},\mathbf{X_{T}})$ for the targets $\mathbf{y_{T}}$ given the corresponding features $\mathbf{X_{T}}$ and the training data $\mathbf{X}, \mathbf{y}$ is defined as a multivariate Gaussian distribution $\mathcal{N}$ with a predictive mean $\mathop{\mathbb{E}}[\mathbf{y_{T}}|\mathbf{X},\mathbf{y},\mathbf{X_{T}}]$ and a predictive variance $\mathop{\mathbb{V}}[\mathbf{y_{T}}|\mathbf{X},\mathbf{y},\mathbf{X_{T}}]$.

The $\mathbf{X} \in \mathcal{R}^{P\times M}$ and $\mathbf{X_{T}}$ $\in \mathcal{R}^{N\times M}$ are the sets of $M$ features for $P$ samples for training and $N$ samples for testing. The $\mathbf{y} \in \mathcal{R}^{P}$ and $\mathbf{y_{T}}$ $\in \mathcal{R}^{N}$ are the target objectives corresponding to the number of samples per dataset respectively. The  $\mathop{\mathbb{E}}[\mathbf{y_{T}}|\mathbf{X},\mathbf{y},\mathbf{X_{T}}]$ is defined in Equation~\ref{eq:m} below as 
\begin{equation}
     \underline{m(\mathbf{X_{T}})\backsim T_i(\mathbf{X_{T}})} +
     k(\mathbf{X_{T}}, \mathbf{X})(k(\mathbf{X},\mathbf{X}) + \sigma^2\mathbf{I})^{-1}(\mathbf{y}-\underline{m(\mathbf{X})\backsim T_i(\mathbf{X})})
     \label{eq:m}
 \end{equation}
 
 \noindent
 and $\mathop{\mathbb{V}}[\mathbf{y_{T}}|\mathbf{X},\mathbf{y},\mathbf{X_{T}}]$ is defined in Equation~\ref{eq:v} below as 
 \begin{equation}
    k(\mathbf{X_{T}}, \mathbf{X_{T}}) - k(\mathbf{X_{T}}, \mathbf{X})(k(\mathbf{X},\mathbf{X}) + \sigma^2\mathbf{I})^{-1}k(\mathbf{X_{T}}, \mathbf{X})^T
    \label{eq:v}
 \end{equation}
\noindent
where the $\sigma^2$ represents the noise amplitude and $\mathbf{I}$ is the identity matrix\footnote{For a detailed derivation please refer to \cite{rasmussen2003gaussian}.}. In the formulas above, GP possesses a set of hyperparameters associated with both the mean function and the choice of the kernel. The hyperparameter values can be found by maximising the marginal likelihood. The optimal hyperparameters are then chosen by observing the likelihood or by cross-validation.

The GP is usually used with an agnostic mean function centred at zero. However, we propose to use the previously developed latency model $T_{i}$, for each 2D convolutional layer $i$ in a CNN, as a mean function $m(.)$ inside the predictive mean to encapsulate the known analytic model of the accelerator into the proposed method. It uses the collected data $\mathbf{X}$, which in this case are the parameters, and the hardware configuration of the accelerator for each convolution, which would normally be used in the standard analytic estimation. By also recording our past measurements from our past implementations $\mathbf{y}$, we can form a training set on which we can learn the nonlinearities that cannot be analytically modelled. The $\mathbf{X_{T}}$ represents the set of test features corresponding to the 2D convolutions for which we would like to estimate their target performance $\mathbf{y_{T}}$, in this case, latency.

Therefore, the advantage of this method in comparison to other machine learning (ML) inspired methods is that it avoids completely relying on the data while estimating the performance. Additionally, this method does not need to extract any features from the data because the features for the estimation are already known and they are the ones used in the standard analytic estimation. Hence, this method reuses previously developed knowledge by incorporating the standard method into the model as the mean function of the GP to anchor the estimate within reliable bounds. By anchoring the estimate, the model is also more interpretable in comparison to purely data-reliant methods which depend completely on the learnt features which are usually not human-readable. Additionally, by specifying the mean function and combining it with the collected data, the proposed method can give a prediction outside the observed data sample without collapsing.

In the next Section, we present the FPGA-based accelerator from which we have collected the data and onto which we have evaluated our proposed method.

\section{Accelerator and Dataset}\label{sec:acc_data}
\subsection{Accelerator's Architecture}

The per-layer latency of an implemented FPGA-based CNN accelerator is characterised according to the standard method into three parts: (1) Loading time for loading the input, (2) Computation time, (3) Storing time for storing the results. 

The input has to be loaded into the on-chip memory only once for the first layer, similarly to the output being stored only once from the on-chip memory to the off-chip memory. The output of intermediate layers is buffered in the on-chip memory. 

The notation is shown in Table \ref{tab:notation} and the size of the weights and input/output for convolution is shown in Table \ref{tab:conv_operations}. Following the standard method, the per-layer latency $T_i$ for a single input is shown in Equations~\ref{eq:loading_time},~\ref{eq:compute_time} and~\ref{eq:store_time} as follows
\begin{enumerate}

  \item Loading time i.e., the time to load the input into the on-chip memory
  \begin{align}
      T_{weights_i} &=   \frac{K_i \times K_i \times F_i \times C_i \times DW}{PF \times M_{CLK} \times S \times M_{EFF}} \nonumber \\ 
      T_{data_i} &= \frac{H_i \times W_i \times C_i \times DW}{PF  \times M_{CLK} \times S \times M_{EFF}} \nonumber \\ 
      T_{load_i} &= T_{weights_i} +T_{data_i}
      \label{eq:loading_time}
  \end{align}

  \item Computation time i.e., the time to compute $PC \times PF$ parallel channels and filters respectively
  
  \begin{equation}
      T_{compute_i} = \frac{F_i \times C_i \times H_i \times W_i \times K_i \times K_i}{PF \times PC \times L_{CLK}} \label{eq:compute_time}
  \end{equation}

  \item Storing time i.e., the time to store the output back to the off-chip memory
  \begin{equation}
      T_{store_i} = \frac{H_{O_i}\times W_{O_i} \times F_i \times DW}{PF  \times M_{CLK} \times S \times M_{EFF}}
      \label{eq:store_time}
  \end{equation}
\end{enumerate} 

Therefore, the time required to process a single 2D convolutional layer can be written as in Equation \ref{eq:exe_time_heuristic} below as
\begin{equation}
T_{i} = 
\begin{cases}
 T_{i=1} &= T_{load_i} + T_{compute_i}  \\
 T_{i\neq 1 \lor N} &= max(T_{weights_i}, T_{compute_i}) \\
 T_{i=N} &= max(T_{weights_i}, T_{compute_i}) + T_{store_i}
\end{cases}
\label{eq:exe_time_heuristic}
\end{equation}

\subsection{Dataset}\label{sec:dataset}

The evaluation dataset comprises of several different configurations of 2D convolutional layers which are the building blocks of three different CNNs, namely SSD \cite{SSD} with 24 2D convolutions, Yolo \cite{Yolo} with 75 2D convolutions and ResNet-50 \cite{He2016DeepRL} with 57 2D convolutions. SSD and Yolo are characteristic for their irregularities, which results in the output being produced at different times, while the ResNet is known for its residual blocks. Each network was trained in 32-bit floating-point representation and then linearly quantised into 8-bit integer representation \cite{fan2018real}. In total giving $P$ training samples $\mathbf{X}$ as 156 and the input feature size $M$ being 15 corresponding to the first 15 parameters in the Table \ref{tab:notation}. The recorded latency per each convolution represents the targets $\mathbf{y}$.

Each network was executed on the implemented accelerator on Intel Arria GX 1150 FPGA. The analysis of the dataset together with the evaluation parameters can be found in Tables \ref{tab:dataset} and \ref{tab:accelerator_parameters}.

\begin{table}
\begin{minipage}{.49\linewidth}
\centering
\caption{Dataset for evaluation.}
\label{tab:dataset}
\begin{tabular}{|l|l|l|l|}
    \hline
    \textbf{Parameter}                       &   \textbf{Min} &   \textbf{Mean} &   \textbf{Max} \\
    \hline
     $H/W$                      &         1 &     42 &       418 \\ \hline
     $H_O/W_O$                    &         1 &     37 &       416 \\ \hline
     $K$                    &         1 &      2 &         7 \\ \hline
     
     $C$                    &         3 &    360 &      2048 \\ \hline
     $F$                 &        64 &    371 &      2048 \\ \hline
     Latency [ms]            &     0.018 &   0.841 &    11.727 \\ \hline
    \end{tabular}
\end{minipage}
\begin{minipage}{.49\linewidth}
\centering
\caption{Evaluation parameters.}
\label{tab:accelerator_parameters}
\begin{tabular}{|l|l|}
        
        \hline
         \textbf{Parameter}     & \textbf{Value} \\ \hline
         $PC$            & 64               \\ \hline
         $PF$              & 64               \\ \hline
         $M_{CLK}$  & 200 MHz                    \\ \hline
         $L_{CLK}$  & 200 MHz                   \\ \hline
         $M_{EFF}$  & 70\%                    \\ \hline
         $S$      & 64-bit               \\ \hline
         $DW$     & 8-bit      \\ \hline
        \end{tabular}
\end{minipage}
\end{table}

\section{Evaluation}\label{sec:eval}
In evaluation, the proposed method is compared with the standard method, including a GP with a zero mean function, a GP with the ANN mean function \cite{gp_dnn}, LR, GTB and ANN. The dataset described in Section \ref{sec:dataset} is being used to evaluate all these methods. 

For a more comprehensive evaluation, leave-one-out cross-validation (LOO\-CV) with respect to the mean absolute error (MAE) is used to compare the estimators. LOOCV is a particular case of leave-$k$-out cross-validation where $k=1$, which means that a model is trained on all samples except one, onto which the performance is then evaluated. In this instance, the performance of the predictor is measured by the absolute error between the prediction and the target value. The error is accumulated for all samples from which the mean is then calculated by dividing the total summed error by the number of samples.

This approach was also used to determine the best hyperparameters for each regressor with respect to the LOOCV MAE. The results, as well as the individual properties and implementation details for the estimators, are summarised in Table \ref{tab:evaluation_results}.
\begin{table}
\centering
\caption{Evaluation of latency estimation for different methods.}
\label{tab:evaluation_results}
\begin{tabular}{|l|r|r|l|} 
\hline
                                                                           \textbf{Methods} & \begin{tabular}[c]{@{}r@{}}\textbf{LOOCV}\\ \textbf{MAE [ms]}~~\end{tabular} & \begin{tabular}[c]{@{}r@{}}\textbf{Implementation}\\ \textbf{and} \\ \textbf{Optimiser}\end{tabular}   & \textbf{Properties}                                                                                                                           \\ 
\hline
Standard method                                                    & 0.450                                                               & None                                    & None                                                                                                                                          \\ 
\hline
 Gaussian process                                                                & 0.521                                                               & GPFlow \cite{GPflow2017} - Adam \cite{kingma2014adam}                           & \begin{tabular}[c]{@{}l@{}}\textit{Mean function:}\\Zero \\\textit{Learning rate:}\\0.001\\\textit{Best kernel:}\\Mat\'ern 3/2\end{tabular}                                   \\ 
\hline
\textbf{Our method}                                                         & \textbf{0.312}                                                               & GPFlow \cite{GPflow2017} - Adam \cite{kingma2014adam}                           & \begin{tabular}[c]{@{}l@{}}\textit{Mean function:}\\ $T_i$\\\textit{Learning rate:}\\0.001\\\textit{Best kernel:}\\Mat\'ern 3/2\end{tabular}                                     \\ 
\hline
\begin{tabular}[c]{@{}l@{}} Gaussian process with\\ Artificial neural network\\ mean function\end{tabular} & 0.692                                                               & GPFlow \cite{GPflow2017} - Adam \cite{kingma2014adam}                           & \begin{tabular}[c]{@{}l@{}}\textit{Mean function:}\\Artificial neural \\ network \\15, 64, 1 nodes and\\ tanh activations\\\textit{Learning rate:}\\0.00001\\\textit{Best kernel:}\\Mat\'ern 3/2\\\end{tabular}  \\ 
\hline
Linear regression                                                                 & 0.450 & sklearn \cite{scikit-learn}                                 & Default                                                                                                                                          \\ 
\hline
Gradiet tree boosting                                                                & 0.607                                          & sklearn \cite{scikit-learn} - AdaBoost \cite{friedman2002stochastic} & \begin{tabular}[c]{@{}l@{}}\textit{Learning rate:}\\0.1\\\textit{Number of trees:}\\10\\\textit{Maximum depth:}\\ 3\end{tabular}                                           \\ 
\hline
Artificial neural network                                                                &1.257                                          & Tensorflow \cite{tensorflow2015-whitepaper} - Adam \cite{kingma2014adam} & \begin{tabular}[c]{@{}l@{}}\textit{Batch size:}\\8\\\textit{Learning rate:}\\0.1\\\textit{Regulariser:}\\L2, 0.001\\\textit{Number of nodes:}\\10,10,1\\\textit{Activations:}\\ReLU\end{tabular}  \\
\hline
\end{tabular}
\end{table}
We considered several hyperparameters for the proposed GP-based method such as the learning rate, ranging from 0.1 to 0.000001 on a logarithmic scale and the kernel, ranging from linear, Gaussian to Mat\'ern kernels \cite{rasmussen2003gaussian} and their combinations. The best parameters were found by a grid search with respect to the LOOCV MAE.

In case of the GP with the ANN mean function, it was necessary to find hyperparameters for the ANN such as the number of nodes in the hidden layers, between 16, 32 and 64 and the number of hidden layers, ranging from 1 to 3. For the activation function, we considered tanh, ReLU and sigmoid. For GTB and ANN, we needed to determine the most influential parameters such as the learning rate, ranging from 0.01 to 0.0001 on a logarithmic scale or for the GTB, the number of trees or the tree depth that was determined by gradual pruning. For the ANN we needed to decide the number of hidden nodes, between [10, 1], [10, 10, 1] and [10, 10, 10, 1] and for the activation function, we again considered tanh, ReLU and sigmoid. The hyperparameters were similarly found through a grid search with respect to the LOOCV MAE. For the standard method and LR, it was not necessary to determine any hyperparameters. 

Overall, the best method proved to be the combination of the standard method and the collected data in the form of the GP with an analytic mean function. In comparison to other approaches, the proposed method achieved approximately a {30.7\%} improvement in LOOCV with respect to MAE decreasing to $0.312\ ms$ in comparison to the second best-performing methods, which were LR and the standard method with $0.450\ ms$ MAE.

The main advantage of the method lays in its implementation simplicity, as it reuses the analytic approximation that is commonly used for DSE, combined with recorded measurements. The method can be improved by recording more measurements and simple fine-tuning of the hyperparameters related to the kernel $k$ or the analytic mean $m$.

A potential limitation of this method stems from the kernel computation which scales with the complexity of  $O(P^3)$, which means that the inference time can be prolonged if there are many training samples. One possible solution to overcome this problem is using k-Means clustering to determine the $k$ most important points that have to be included in the kernel. Nevertheless, the inference time is much less than the time needed for synthesis and then running the design on hardware.

\section{Conclusion and Future Work}\label{sec:conclusion}

In this paper, we proposed an accurate method for estimating the performance of an field-programmable gate array-based accelerator for convolutional neural networks and compared it with the standard method and variations of the Gaussian process, linear regression, gradient tree boosting and an artificial neural network. The evaluation demonstrated that the innovative Gaussian process paired with the domain-specific knowledge and collected data can provide an approximately {30.7\%} accuracy improvement with respect to the standard method or the linear regression. 

In the proposed method, users need to decide what are the relevant software/hardware features $M$ together with an analytic approximation for the modelled performance that will be used as the mean function $m(.)$ in the Gaussian process. Afterwards, they need to supply the profiling data for training $\mathbf{X}, \mathbf{y}$, $\mathbf{X}$ is the feature matrix and $\mathbf{y}$ are the targets, in this case, the per-layer latency. In the end, the user needs to decide what is going to be the best kernel $k(.,.)$ and use it to train the Gaussian process to obtain the best values for the hyperparameters.\footnoteref{note1}

In the future, we will validate the method on more configurations on different hardware boards. Furthermore, we will formulate similar analytic approximations for other potential objectives, for example, the resource usage or power consumption and use them as priors for estimating these objectives through our proposed Gaussian process-based method.

\subsubsection{Acknowledgments.} We thank Yann Herklotz, Alexander Montgomerie-Corcoran and ARC'20 reviewers for insightful suggestions.

\bibliographystyle{splncs04}
\bibliography{sample}

\end{document}